\documentclass[letterpaper,fleqn,usenatbib]{mnras}

\usepackage{enumerate}

\def\eg{{\it e.g.}}
\def\etal{{\it et al.}}

\def\ie{{\it i.e.}}

\def\pmb#1{\setbox0=\hbox{$#1$}%
  \kern-0.25em\copy0\kern-\wd0
  \kern.05em\copy0\kern-\wd0
  \kern-0.025em\raise.0433em\box0}
\def\spmb#1{\setbox1=\hbox{${\scriptstyle #1}$}%
  \kern-0.25em\copy1\kern-\wd1
  \kern.05em\copy1\kern-\wd1
  \kern-0.025em\raise.0433em\box1}

\long\def\Ignore#1{\relax}
\usepackage{color}
\definecolor{red}{rgb}{0.7,0.1,0.1}
\definecolor{blue}{rgb}{0.2,0.2,0.8}
\definecolor{green}{rgb}{0.1,0.6,0.1}

\usepackage{graphicx}	

\title[Spiral instabilities]{Spiral instabilities: little interaction with a live halo}

\author[Sellwood]
          {J. A. Sellwood,$^{1}$\thanks{E-mail:sellwood@as.arizona.edu}
\\
$^1$Steward Observatory, University of Arizona, 933 N Cherry Ave,
Tucson AZ 85722, USA}

\pubyear{2021}

\begin{document}
\label{firstpage}
\pagerange{\pageref{firstpage}--\pageref{lastpage}}
\maketitle

\begin{abstract}
In order to address the question of whether spiral disturbances in
galaxy discs are gravitationally coupled to the halo, we conduct
simulations of idealized models of disc galaxies.  We compare growth
rates of spiral instabilities in identical mass models in which the
halo is held rigid or is represented by particles drawn from an
equilibrium distribution function.  We examine cases of radial and
azimuthal bias in the halo velocity ellipsoid in one of our models,
and an isotropic velocity distribution in both.  We find at most
marginal evidence for an enhanced growth rate of spiral modes caused
by a halo supporting response.  We also find evidence for very mild
dynamical friction between the spiral disturbance and the halo.  We
offer an explanation to account for the different behaviour between
spiral modes and bar modes, since earlier work had found that bar
instabilities became significantly more vigorous when a responsive
halo was substituted for an equivalent rigid mass distribution.  The
barely significant differences found here justify the usual
simplifying approximation of a rigid halo made in studies of spiral
instabilities in galaxies.

\end{abstract}



\section{Introduction}
\label{sec.intro}
Theoretical studies of spiral instabilities in galaxies
\citep[\eg][]{To81, BLLT, SK91} have almost always assumed that bulge
and halo components can be treated as inert matter that contributes to
the central attraction, but does not otherwise affect the dynamics of
the disc component.  This assumption is often made in simulations also
\citep[\eg][]{GKC12, DVH13, SC14, Baba15}.  Exceptionally, \citet{Mark76}
argued that spiral modes would be amplified by the loss of angular
momentum to the halo, which he predicted would take place mostly at
corotation of the spiral mode.  While many simulations of discs
embedded in live halos have been performed in subsequent years, there
does not appear to have been a quantitative test of Mark's prediction.

It has long been established that strong bars experience dynamical
friction from responsive halos \citep[\eg][]{Sell80, Wein85, DS00}.
It has also been found that bar instabilites are more vigorous in
responsive halos \citep[\eg][]{Athan02, SaNa13, BeSe16} than in rigid
ones, and that the more rapid growth begins at tiny amplitude
\citep{Sell16}.  This last result suggests that a rigid halo might
also be an inadequate approximation for studies of spiral
instabilities.

This paper therefore reports a suite of idealized simulations of discs
in halos to determine the effect on spiral modes of subsitituting a
responsive halo for a rigid one.

\section{Technique}
\label{sec.methods}
When using simulations to make quantitative measurements of the growth
rates of instabilities, it is important to set up the initial model
with considerable care in order that it be close to a settled
equilibrium.  Even quite mild adjustments to the initial model will
mask, and perhaps seed, early growth of instabilities, and the
instability will be that of the relaxed model and not that of the mass
distribution originally intended.  Equilibrium models of a disc
embedded in a halo present a particular challenge, since few
equilibrium distribution functions (DFs) are known for the halo
component when a massive disc is present.  In this paper we present
two separate equilibrium models and provoke a spiral instability in
the disc component of each.  We then compare the evolution of the
instability in both the fully-self consistent models having reponsive
halos and cases in which the halos are frozen.

\subsection{Numerical method}
The particles in our simulations move in a 3D volume that is spanned
by two separate grids; a cylindical polar mesh and a much larger
spherical grid. The gravitational field is calculated at grid points
and interpolated to the position of each particle.  The disc particles
are initially assigned to the polar grid, while the spherical grid is
used for the bulge and halo particles.  Naturally, all particles are
attracted by all others at every step.  A full description of our
numerical procedures is given in the on-line manual \citep{Se14} and
the code itself is available for download.

Tables~\ref{tab.PKpars} and \ref{tab.DBHpars} give the values of the
numerical parameters for most of the simulations presented in this
paper.  Since the gravitational field is a convolution of the mass
density with a Green function that is most efficiently computed by
Fourier transforms, it is easy to restrict the sectoral harmonics that
contribute to the field when using a polar grid.  In most simulations
we report here, non-axisymmetric forces arising from the particles are
confined to just the $m=3$ sectoral harmonic for the PK model
(\S\ref{sec.PKmodel}) and $m =2$ for the DBH model
(\S\ref{sec.DBHmodel}).  Aspherical components of forces computed via
the spherical grid are similarly confined to the spherical harmonic $l
= 3$ or 2.

\subsection{Other details}
We make measurements of the development of non-axisymmetric
disturbances from the azimuthal Fourier coefficients of the mass
distribution on the polar grid, generally combining them to report a
weighted average of the relative amplitude over a limited radial
range.  We also compute logarithmic spiral transforms of the disc
particle distribution, which is defined for particles of unequal mass
as
\begin{equation}
A(m,\tan\gamma,t) = {\sum_{j=1}^N \mu_j\exp[im(\phi_j + \tan\gamma \ln R_j)]
\over \sum_{j=1}^N \mu_j},
\label{eq.logspi}
\end{equation}
where $(R_j,\phi_j)$ are the cylindrical polar coordinates of the
$j$-th particle of mass $\mu_j$ at time $t$, and $\gamma$, the
complement to the spiral pitch angle, is the angle between the radius
vector and the tangent to an $m$-arm logarithmic spiral, with positive
values for trailing spirals.

In order to measure mode frequencies, including growth rates, we fit
modes to these data using the procedure described by \citet{SA86}.
The perturbed surface density of a mode is the real part of
\begin{equation}
\delta\Sigma(R,\phi,t) = {\cal A}_m(R)e^{i(m\phi - \omega t)},
\label{eq.mode}
\end{equation}
where the frequency $\omega = m\Omega_p + i\beta$ is complex with
$\beta$ being the growth rate.  The complex function ${\cal A}_m(R)$,
which is independent of time, describes the radial variation of
amplitude and phase of the mode.

\citet{SA86} showed that the mode frequency, especialy the growth
rate, could be estimated more precisely from quiet start simulations
in which noise was artificially suppressed by placing particles around
rings and filtering out the higher sectoral harmonics, allowing a
longer period of exponential growth.  As this trick is more difficult
to accomplish in 3D multi-component models, we here beat down shot
noise simply by employing large numbers of particles.

\begin{table}
\caption{Numerical parameters for the PK model}
\label{tab.PKpars}
\begin{tabular}{@{}ll}
Polar grid size & 85 $\times$ 128 $\times$ 125 \\
Grid scaling & $a= 10$ grid units \\
Vertical spacing & $\delta z = 0.02a$ \\
Active sectoral harmonics & $m = 0$, 3 \\
Softening length & $a/10$ \\
Spherical grid & 201 shells \\
Active spherical harmonics & $l = 0$, 3 \\
Disc mass fraction & $f_d = 1/3$ \\
Number of disc particles & $10^8$ \\
Number of halo particles & $10^8$ \\
Basic time-step & $(a^3/GM)^{1/2}/40$ \\
Time step zones & 3 \\
\end{tabular}
\end{table}

\section{Plummer-Kuzmin model}
\label{sec.PKmodel}
A composite disc and halo model that is very nearly in equilibrium can
be constructed using a Kuzmin disc embedded within a Plummer sphere.
The potential of an axisymmetric Kuzmin disc is
\begin{equation}
\Phi_{\rm K}(R,z) = -G M \left[R^2 + (a + |z|)^2\right]^{-1/2},
\label{eq.phitot}
\end{equation}
\citep[][eq.~2.68a]{BT08}, where $M$ is the mass of the disc and $a$
is a length scale.  That of a Plummer sphere, also of mass $M$ and
core radius $a$, is $\Phi_{\rm P}(r) = -GM[r^2 + a^2]^{-1/2}$, which
is identical in the $z=0$ plane and differs only slightly for $|z| >
0$.  Thus, a composite model of two superposed concentric components
having equal length scales $a$, with disc mass $f_dM$ and sphere mass
$(1-f_d)M$, will have the same potential in the mid-plane.  We
describe this as the PK model, and it is natural to adopt units such
that $G = M = a = 1$.  Though hardly a realistic galaxy model, it has
the following advantages for our study.

\citet{Kaln76} gave a family of equilibrium DFs having a parameter
$m_K$ that determines the degree of random motion in the razor-thin
Kuzmin disc model.  Also \citet{Dejo87} derived a family of DFs having
a parameter $q$ that determines the shape of the velocity ellipsoid in
the Plummer sphere.  Neither of these DFs would yield perfect
equilibrium in a composite model, however. The Plummer sphere would be
affected by the slight difference of the disc contribution to the
potential away from the mid-plane, which we find is small enough to
ignore.  More significantly, the disc needs to be thickened to avoid
local disc instabilities.  A disc lacking random motion is Jeans
unstable \citep{To64}, while a razor-thin disc having a finite radial
velocity dispersion suffers from buckling instabilities \citep{To66,
  Arak85}.  Since thickening the disc weakens the central attraction
in the mid-plane, as does gravity softening, radial balance requires a
compensating additional central force that is the difference between
that expected from the total mid-plane potential and the numerically
determined attraction from the particles in the simulation.  The
radial variation of the extra central force needed in the mid-plane is
computed at the start, and is applied as a spherically symmetric term
that is held fixed throughout the simulation.  It can be thought of as
an additional rigid halo, although the implied mass is just a few
percent of that of the disc.

The principal advantage of pursuing these oversimplified galaxy models
is that we can employ anisotropic DFs of the Plummer sphere in our
experiments.  We wish to determine the influence of a responsive halo
on the spiral instabilities within the disc, and this model enables us
to vary the shape of halo velocity ellipsoid from radial bais, through
isotropy to azimuthal bias, as was also exploited by \citet{Sell16} in
his study of bar instabilities.  Furthermore, we are able to set up
good equilibria for both the disc and halo by drawing particles from
the known DFs, using the method described by \citet{DS00}.  We use the
1D Jeans equation to set up the vertical balance of the disc
particles, which yields an acceptable equilibrium when the radial
velocity dispersion is modest.  The resulting model is very close to
overall equilibrium, and the mass profiles of both disc and halo
remain indistinguishable from the initial set up during the entire
linear growth phase of the instability.

Since the bar instability in a disc is made even more vigorous by a
live halo, spiral instabilities would be overwhelmed by the bar mode
unless the disc mass fraction were very low, and then the spirals
would be weak and multi-armed.  We therefore restrict non-axisymmetric
disturbance forces in these simulations to the $m = 3$ sectoral
harmonic in order to inhibit the bar mode, while allowing a single
spiral mode.

We provoke a spiral instability at a predictable radius by creating a
groove in the angular momentum profile of the disc \citep{SK91}.  In
this case, we multiply the DF given by \citet{Kaln76}, which is a
function of the two classical integrals $E$ and $L_z$, by the
Lorentzian factor
\begin{equation}
\left[ 1 - {D  w_j^2 \over ( L_z - L_z^* )^2 + w_j^2 } \right],
\label{eq.groove}
\end{equation}
and select equal mass disc particles from this modified DF.  In
expression (\ref{eq.groove}), $D$ is the relative depth of the
groove, $w_j$ is the width parameter, and $L_z^*$ is the angular
momentum of the groove centre.  Note that this is a groove in the
distribution of guiding centres and epicyclic blurring implies a
shallower and broader dent in the surface density profile.

For our simulations, we choose $f_d = 1/3$ and the disc DF parameter
$m_K = 30$, which yields $Q \simeq 2$ in the disc centre and
decreasing gently to $Q\simeq1.2$ in the outer disc.  The groove
parameters are: $L_z^* = 1.4$, $w_j= 0.04$ in our adopted units, and
$D=0.4$.  The halo DF parameters in the three models with live
halos are $q=0$ for an isotropic velocity distribution, $q=2$ for
maximum radial bias, and $q=-15$ for a strong azimuthal bias.  Note
that in all cases the central parts of the halo are closely isotropic,
while anisotropies increase with radius beyond the core.

\begin{figure}
\includegraphics[width=.95\hsize,angle=0]{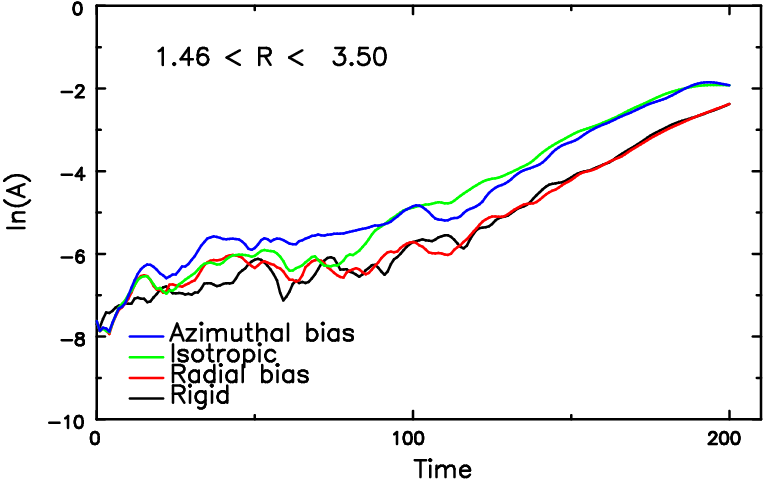}
\begin{center}
\includegraphics[width=.92\hsize,angle=0]{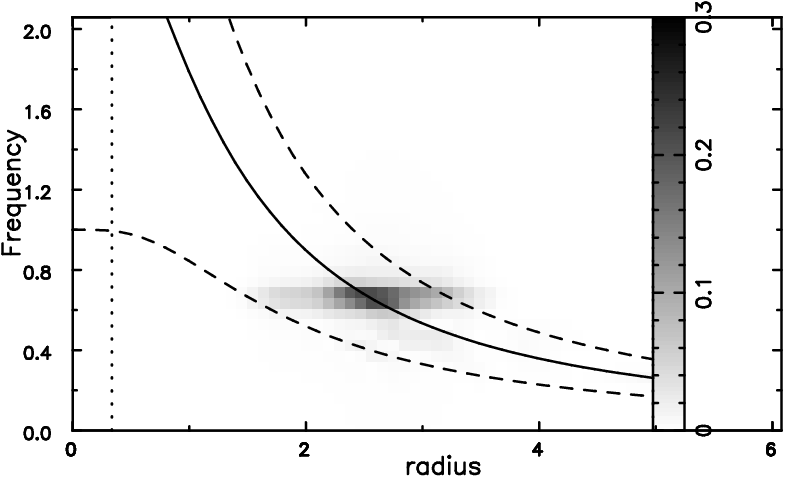}
\includegraphics[width=.8\hsize,angle=0]{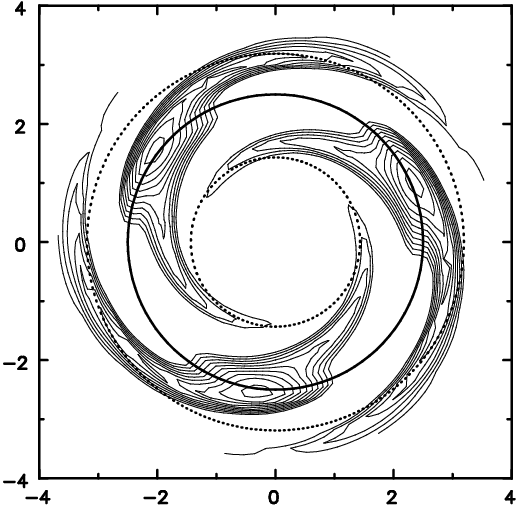}
\end{center}
\caption{Top panel: The weighted average of the $m=3$ relative
  overdensity in the indicated radial range over which the spiral mode
  is strongest.  The four separate simulations differ only in the halo
  DF as indicated by the line colours, and the black line is for a rigid
  halo.  Middle panel: Power spectrum of the $m=3$ component from the
  isotropic halo model over the time interval $61 \leq t \leq 180$
  indicating a single dominant mode.  The solid curve marks
  $m\Omega_c$ and the dashed curves $m\Omega_c \pm \kappa$.  Bottom
  panel: The mode fitted to the data from the isotropic halo model
  over the time interval $80 \leq t \leq 180$.  The corotation radius
  is marked by the full-drawn circle and the Lindblad resonances by
  the dotted circles.}
\label{fig.PK-res}
\end{figure}

\subsection{Results}
\label{sec.results}
We have conducted four simulations of the PK model, each employing
100M particles in the disc component.  They differ in the shape of
velocity ellipsoid in the live halo component, which also has 100M
particles and in one case the halo was replaced by a rigid
unresponsive mass of the same density.  The groove provoked an $m=3$
spiral instability in all four cases.  The power spectrum in the
middle panel of Figure \ref{fig.PK-res} indicates that the isotropic
halo model supported a single dominant mode with corotation at
$R\simeq 2.6$; power spectra from the other three simulations were
very similar.  The upper panel shows the growth of the $m=3$
disturbance amplitude in the disc component in each PK model.  The
instability took slightly different lengths of time to emerge from the
noise, but over the interval $100 \leq t \leq 180$ the slope of each
line in this log-linear plot, which is the growth rate of the mode, is
remarkably similar.  This finding is already in strong contrast with
the behaviour reported for bar instabilities by \citet{Sell16}, where
growth rates in models with different halo velocity ellipsoids were
between 2 and 5 times greater than the bar mode in a disc with rigid
halo.

\begin{table}
\caption{Mode frequencies fitted to the PK models}
\label{tab.PKmodes}
\begin{tabular}{@{}lcc}
Halo  & $m\Omega_p$ & $\beta$ \\
\hline
Rigid &  $0.663\pm0.002$ &  $0.039\pm0.004$ \\
Isotropic & $0.659\pm0.001$ & $0.040\pm0.002$ \\
Azimuthal bias &  $0.653\pm0.002$ & $0.045\pm0.004$ \\
Radial bias & $0.660\pm0.001$ & $0.041\pm0.002$ \\
\end{tabular}
\end{table}

We have estimated the mode frequencies in all four models by the
method described in \citet{SA86}, fitting eq.~(\ref{eq.mode}) to data
from the simulations over selected time ranges.  We fit both the
disturbance density and logarithmic spiral transforms, selecting
differing time ranges, radial ranges, and numbers of values of
$\tan\gamma$.  We use these different fits to obtain estimates of the
uncertainties in the fitted frequencies.  The best fit values are
presented in Table~\ref{tab.PKmodes}, with the quoted uncertainties
bracketing the full range of credible fitted values in each case.  The
mode shape drawn in bottom panel of Figure \ref{fig.PK-res} is the
real part of best fit function ${\cal A}_m(R)$ (eq.~\ref{eq.mode}),
and the contours are of only the positive relative overdensity; the
principal resonances are marked by circles.

Only the model with the strong azimuthal bias has a higher growth rate
than the others, though barely so.  This finding is supported by the
blue line in the upper panel of Figure \ref{fig.PK-res}, which may
have a visibly steeper slope over the time interval $100 < t <180$
than the other three cases.  But it seems that the mode grows at
essentially the same rate in the rigid halo as in the isotropic and
radially biased halos, and scarcely any faster in the azimuthally
biased halo.

A supporting response from the halo would most likely come from those
orbits that are near circular, directly rotating, and oriented close
to the disc plane, since they would effectively augment the disc, as
\citet{Sell16} argued for bar modes.  As the number of such orbits
should be greatest when the halo velocity ellipsoid has a strong
azimuthal bias, a significant supporting response seems physically
reasonable, but even in this case the growth rate is enhanced by no
more than $\sim10$\%.  However, the trefoil spiral in these models may
not induce as strong a halo supporting response as would a
bi-symmetric spiral, as has been established for bar modes.

\begin{figure}
\includegraphics[width=.9\hsize,angle=0]{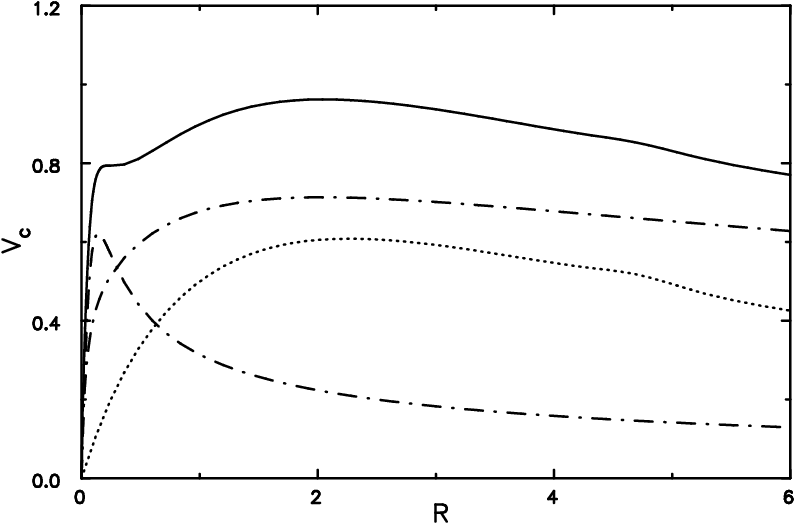}
\caption{The initial rotation curve of the disc-bulge-halo model
  (solid curve) measured from the particles.  The disc contribution is
  given by the dotted line, while the dot-dash lines indicate those of
  the bulge and compressed halo.}
\label{fig.DBH-rc}
\end{figure}

\section{Disc-bulge-halo model}
\label{sec.DBHmodel}
In order to study the behaviour in a somewhat more realistic model,
and to extend it to bisymmetric spirals, we adopt an exponential disc
model embedded within a halo and a central bulge.  The bulge is
required both to create the steeper inner rise in the rotation curve
typical of a large spiral galaxy, and also to prevent bar
instabilities \citep{To81} while allowing $m=2$ spirals.

\begin{itemize}
\item The exponential disc has the surface density
\begin{equation}
\Sigma(R) = {M_d \over 2\pi R_d^2} \exp(-R/R_d),
\end{equation}
which we taper to zero over the radial range $4.5 < R/R_d < 5$, and
$M_d$ is the mass of the notional infinite disc.  The vertical density
profile of the disc is Gaussian with a scale $0.1R_d$.  

\item We use a dense Plummer sphere for the bulge, which has a mass of
  $0.1M_d$ and core radius $a=0.1R_d$, and adopt the isotropic DF
  given by \citet{Dejo87}.  The bulge is dense enough that it
  dominates the central attraction in the inner part of our galaxy
  model, and the analytic DF is close to equilibrium despite the
  presence of the disc and halo.

\item We employ a \citet{Hern90} model for the halo that has the
  density profile
\begin{equation}
\rho(r) = {M_h b \over 2\pi r ( b + r )^3},
\end{equation}
where $M_h$ is the total mass integrated to infinity, and $b$ is a
length scale.  We choose $M_h = 5M_d$ and $b=4R_d$.  Naturally, the
isotropic DF that Hernquist derived for this isolated mass
distribution would not be in equilibrium when the disc and bulge are
added, so we use the adiabatic compression procedure described by
\citet{SM05} that starts from the isotropic DF given by Hernquist and
uses the invariance of both the radial and azimuthal actions to
compute a revised density profile and DF as extra mass is inserted.
The revised DF has a slight radial bias.  We also apply an outer
cutoff to the selected particles that excludes any with enough energy
ever to reach $r>10b$, which causes the halo density to taper smoothly
to zero at that radius.
\end{itemize}

\begin{table}
\caption{Numerical parameters for DBH model}
\label{tab.DBHpars}
\begin{tabular}{@{}ll}
Polar grid size & 85 $\times$ 128 $\times$ 125 \\
Grid scaling & $R_d= 10$ grid units \\
Vertical spacing & $\delta z = 0.02R_d$ \\
Active sectoral harmonics & $m = 0$, 2 \\
Softening length & $R_d/10$ \\
Spherical grid & 501 shells \\
Active spherical harmonics & $l = 0$, 2 \\
Number of disc particles & $10^8$ \\
Number of halo particles & $10^8$ \\
Number of bulge particles & $10^7$ \\
Basic time-step & $(R_d^3/GM)^{1/2}/320$ \\
Time step zones & 6 \\
\end{tabular}
\end{table}

The resulting rotation curve of this model is illustrated in
Figure~\ref{fig.DBH-rc}, where we have adopted units such that $G =
M_d = R_d = 1$.  We also adopt a constant value of $Q=1.5$ at all
radii to determine the radial velocity dispersion of the disc
particles.  Although the disc is quite massive, the high value of the
epicyclic frequency, $\kappa$, near the centre in particular implies
disc random velocities are modest everywhere, and the Jeans equations
in the epicyclic approximation \citep{BT08} yield an excellent
equilibrium.

\begin{figure}
\includegraphics[width=.95\hsize,angle=0]{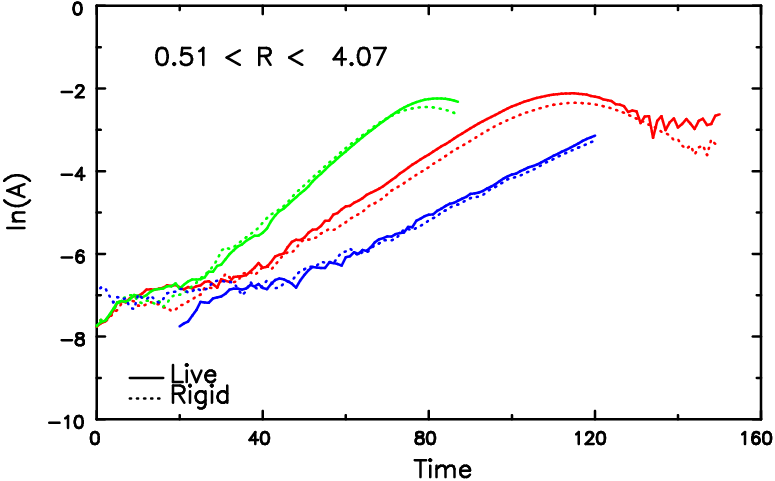}
\begin{center}
\includegraphics[width=.92\hsize,angle=0]{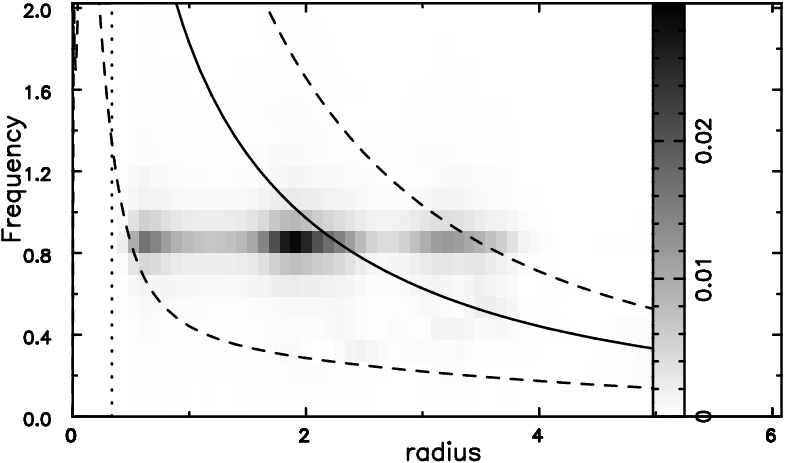}
\includegraphics[width=.8\hsize,angle=0]{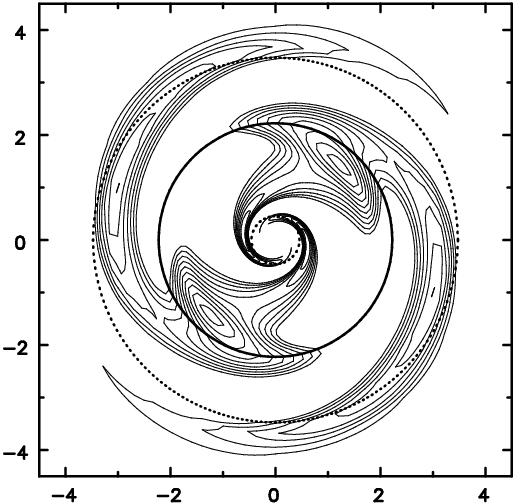}
\end{center}
\caption{Top panel: The weighted average of the $m=2$ relative
  overdensity in the indicated radial range over which the spiral mode
  is strongest.  The six simulations of this DBH model have rigid
  (dotted lines) and live halos (solid lines) and $L_z^* = 2$ (red),
  $L_z^* = 1.5$ (green), and $L_z^* = 2.5$ (blue). The blue lines have
  been shifted horizontally to show the similarity of slopes without
  overlapping with the others.  Middle panel: Power spectrum of the
  $m=3$ component from the live halo simulation with $L_z^* = 2$ over
  the time interval $30 \leq t \leq 80$ indicating a single dominant
  mode.  Bottom panel: The mode fitted to the data from the live halo
  model over the time interval $30 \leq t \leq 90$.  The lines in the
  middle panel and the circles in the bottom panel have the same
  meanings as in Fig.~\ref{fig.PK-res}.}
\label{fig.DBH-res}
\end{figure}

Again we wish to seed a spiral instability at a predictable radius,
but require a different strategy in this case because we do not select
disc particles from a DF.  Instead we create a groove in the disc
surface density profile by multiplying the mass of every disc particle
by the same expression as above (eq.~\ref{eq.groove}).  As before,
this sharp groove in angular momentum is blurred by epicycle motions.

We run three cases placing the groove centre at $L_z^* = 1.5$, 2, and
2.5.  In each case we choose, $w_j= 0.04$ in our adopted units, and
$D=0.6$, and run both live and rigid halos and bulges.  We employ 100M
particles in each of the disc and halo and 10M in the bulge.  The
other numerical parameters are given in Table~\ref{tab.DBHpars}.

\begin{table}
\caption{Mode frequencies fitted to the DBH models}
\label{tab.DBHmodes}
\begin{tabular}{@{}lccc}
Halo & $L_z^*$ & $m\Omega_p$ & $\beta$ \\
\hline
Rigid & 1.5 &  $1.109\pm0.001$ &  $0.093\pm0.001$ \\
Live & 1.5 & $1.108\pm0.004$ & $0.095\pm0.001$ \\
Rigid & 2 &  $0.855\pm0.001$ &  $0.065\pm0.002$ \\
Live & 2 & $0.841\pm0.001$ & $0.069\pm0.003$ \\
Rigid & 2.5 &  $0.674\pm0.001$ &  $0.049\pm0.001$ \\
Live & 2.5 & $0.667\pm0.003$ & $0.051\pm0.001$ \\
\end{tabular}
\end{table}

\subsection{Results}
The middle panel of Figure~\ref{fig.DBH-res} gives the power spectrum
in the model with the $L_z^*=2$ and live halo while the upper panel
indicates the growth of the $m=2$ disturbance in all six simulations
with this galaxy model; those having a responsive halo and bulge and
marked with solid lines and those for which both spherical components
are rigid have dotted lines.  As for the PK models, differences
between the live and rigid halos for all three diferent groove radii
are tiny -- the bi-symmetric spiral mode appears to grow at roughly
the same rate in the live as in the rigid halo in each case.
Naturally, the mode pattern speed and growth rate decrease as the
groove radius is increased because the dynamical clock runs more
slowly.  The fitted frequencies in the six runs are listed in
Table~\ref{tab.DBHmodes}, from which it may be seen that the fitted
growth rates in the live halo runs are fractionally higher than in the
rigid halo cases, but the differences are all within the estimated
uncertainties.

The mode shape for the $L_z^*=2$ case is illustrated in the bottom
panel of Figure \ref{fig.DBH-res}.  The pronounced kink across
corotation reflects the mechanism of the groove mode in a cool disc
\citep{SK91}.  The mode shape is very similar for the rigid halo case
and, apart from a different spatial scale, for the other two groove
radii.

\begin{figure}
\includegraphics[width=.95\hsize,angle=0]{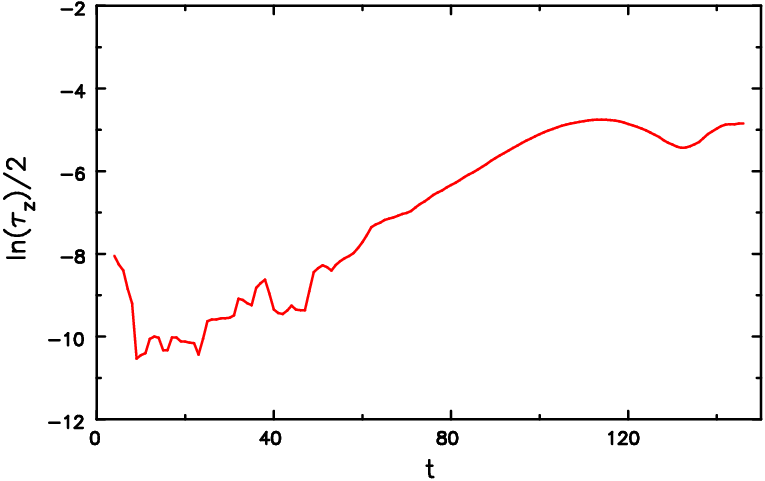}
\caption{The time evolution of the square-root of the torque between
  the disc and halo for the $L_z^*=2 case$.  Notice the quantitative
  scaling similarity between the solid red line in the top panel of
  Fig.~\ref{fig.DBH-res} and that in this figure, indicating that the
  torque varies as the square of the mode amplitude.}
\label{fig.DBH-torque}
\end{figure}

The live halo causes other differences, apart from a marginal increase
in growth rate.  For example in the $L_z^*=2$ case, the saturation
amplitude of the mode is some 20\% higher in the live halo than in the
rigid.  Also a small fraction, $\sim 0.2$\%, of the disc angular
momentum is taken up by the halo, and a tiny fraction, $\sim 0.004$\%,
by the bulge.  Though always small, the torque on the halo
(Figure~\ref{fig.DBH-torque}) increases approximately as the square of
the mode amplitude, \ie\ at twice the growth rate, until it saturates,
which is the scaling expected in dynamical friction as a perturber is
decelerated by its wake \citep{TW84, Sell06, BT08}.

%

It is noteworthy that we could not find any evidence for a bar
instability in this model, in agreement with the linear theory
prediction for a model having a dense centre \citep{To81}.  However,
inner Lindblad resonance damping of disturbances is a prediction of
small amplitude perturbation theory and the strong bisymmetric spiral
extends over a broad radial range, yet it still did not lead to
non-linear trapping into a bar in the inner disc, at least for the
duration of the simulations.

\section{Discussion}
It may seem strange that bar instabilities in responsive halos grow
several times faster than in the equivalent rigid halo \citep{Athan02,
  Sell16}, while a live halo has a tiny effect on even
bisymmetric spiral modes.  Note that this difference appears in the
linear growth phase of the instabilities and therefore the fact that
bar modes may ultimately rise to larger amplitude than do spirals
cannot account for this particular discrepancy.

Another significant difference between the two cases is the second
order dependence of the torque from the spiral instability on the
halo, which contrasts with the first order torque reported by
\citet[][his fig.~2]{Sell16} for the strong halo enhancement of bar
instabilities.

The self-consistency requirement of a linear instability is that the
disturbance density results from orbit deflections caused by the
disturbance potential.  Thus, when a live halo has a large effect on
the growth rate of the bar mode, its supporting response must be an
integral part of the instability of the disc-plus-halo system.
Furthermore, the linear growth of the disc-halo torque indicates that
the instability is driven, in part at least, by angular momentum
loss from the disc.

In contrast, we report here that a live halo has little effect on
spiral modes, indicating that the halo takes no significant part in
this type of instability in the disc-halo system.  Dynamical friction
on the disc disturbance as it moves through the non-rotating halo
is inevitable, but is weak and second order.

So why do we find a negligible halo response to spiral modes, but had
previously reported an enthusiastic supporting response to bar modes?
We suggest the following reason: \citet{LB79} demonstrated that orbits
of arbitrary eccentricity that are subject to the gravitational
influence of a weak bar generally will be repelled from alignment with
the bar {\em except} in the inner parts of galaxy models having a
quasi-uniform core, \ie\ where the rotation curve rises roughly
linearly.  The bar unstable models simulated by both \citet{Athan02}
and by \citet{Sell16} had slowly rising rotation curves, and therefore
the prograde halo orbits near the disc plane in the inner parts of
those models would be attracted into co-alignment with the growing bar
perturbation in the disc, creating a strong supporting response from
the halo that enhances the growth rate.  Where the rotation curve is
flat or declining, the response of higher angular momentum halo
particles does not reinforce the perturbation, and disturbances in the
disc are little affected by a responsive halo.  Thus an aligning halo
response for the bisymmetric spiral in the DBH model would not be
expected because the central bulge eliminates an extensive region
where the potential is quasi-harmonic.  We have reworked Lynden-Bell's
analysis for the trefoil spiral in the PK model, finding that the key
gradient $\partial \Omega_i /\partial L_z|_{J_f}$ indicates
anti-alignment in the region of corotation for the groove mode, and
therefore no supporting response from halo orbits of any eccentricity.
Here $\Omega_i = \Omega_\phi - \Omega_R/3$ and the fast action $J_f =
L_z/3 + J_R$; see \citet{LB79} for a fuller explanation.  This
argument would seem to account both for previously published results
for the bar mode and for the contrasting finding for spiral modes in
this paper.

\section{Conclusions}
\label{sec.concl}
We have constructed two equilibrium models of moderately heavy discs
embedded in live halos.  We created grooves in both discs in order to
provoke strong spiral instabilities and have compared the growth rates
of the spiral modes in simulations that employed both rigid and live
halos.  We cannot exclude that the spiral growth rate was increased by
the halo response in both models; there was a hint of a higher growth
rate for the $m=2$ spiral mode for all three grooves in the live halo
case in the DBH model and a slightly more significant enhanced growth
rate for the $m=3$ mode in the PK model when the halo velocity
ellipsoid was strongly azimuthally biased, but the growth rate
differences were small.  Since halos in real galaxies are not expected
to have azimuthally biased velocity ellipsoids, this mild boost to
spiral growth rates, if real, is unlikely to be of relevance to the
development of spirals in galaxies.

We did find evidence for additional differences between the live and
rigid halos in the DBH models.  The halo was weakly torqued by
dynamical friction from the spiral mode, but the loss of angular
momentum from the disc was only 0.2\% by the end of the simulation
and much less during the linear growth phase.  This exchange may
have allowed the final spiral amplitude to be about 20\% greater
in the live halo case.

The results reported here justify the usual assumption made in
theoretical work and in simulations that a frozen halo is an adequate
approximation when considering the dynamics of spiral formation.

\section*{Acknowledgements}
The author thanks Scott Tremaine for some very helpful comments on a
draft of this paper, Ray Carlberg for useful correspondence, and an
anonymous referee for a careful reading of the paper.  JAS
acknowledges the continuing hospitality of Steward Observatory.

\section*{Data availability}
The data from the simulations reported here can be made available
on request.  The simulation code can be downloaded from
{\tt http://www.physics.rutgers.edu/galaxy}


\bsp	
\label{lastpage}

\begin{thebibliography}{99}
\def\skip#1{ \etal\ }
\def\PhD{PhD thesis.}
\def\rmp{Rev. Mod. Phys.}
\def\rpp{Rep. Prog. Phys.}

\bibitem[Araki(1985)]{Arak85}
Araki, S. 1985, \PhD, MIT.

\bibitem[Athanassoula(2002)]{Athan02}
Athanassoula, E. 2002, \apjl, {\bf 569}, L83

\bibitem[Baba(2015)]{Baba15}
Baba, J. 2015, \mnras, {\bf 454}, 2954

\bibitem[Berrier \& Sellwood(2016)]{BeSe16}
Berrier, J. \& Sellwood, J. A. 2016, \apj, {\bf 831}, 65

\bibitem[Bertin \etal(1989)]{BLLT}
Bertin, G., Lin, C. C., Lowe, S. A. \& Thurstans, R. P. 1989, \apj, {\bf 338}, 104

\bibitem[Binney \& Tremaine(2008)]{BT08}
Binney J. \& Tremaine S. 2008, \textit{Galactic Dynamics} 2nd ed. (Princeton University Press, Princeton NJ)

\Ignore{\bibitem[Chequers \& Widrow(2017)]{CW17}
Chequers, M. H. \& Widrow, L. M. 2017, \mnras, {\bf 472}, 2751}

\bibitem[Debattista \& Sellwood(2000)]{DS00}
Debattista, V. P. \& Sellwood, J. A. 2000, \apj, {\bf 543}, 704

\bibitem[Dejonghe(1987)]{Dejo87}
Dejonghe, H. 1987, \mnras, {\bf 224}, 13

\bibitem[D'Onghia \etal(2013)]{DVH13}
D'Onghia, E., Vogelsberger, M. \& Hernquist, L. 2013, \apj, {\bf 766}, 34

\bibitem[Grand \etal(2012)]{GKC12}
Grand, R. J. J., Kawata, D. \& Cropper, M.  2012, \mnras, {\bf 421}, 1529

\bibitem[Hernquist(1990)]{Hern90}
Hernquist, L. 1990, \apj, {\bf 356}, 359

\bibitem[Kalnajs(1976)]{Kaln76}
Kalnajs, A. J. 1976, \apj, {\bf 205}, 751

\bibitem[Lynden-Bell(1979)]{LB79}
Lynden-Bell, D. 1979, \mnras, {\bf 187}, 101

\bibitem[Mark(1976)]{Mark76}
Mark, J. W-K. 1976, \apj, {\bf 206}, 418

\bibitem[Saha \& Naab(2013)]{SaNa13}
Saha, K. \& Naab, T. 2013, \mnras, {\bf 434}, 1287

\bibitem[Sellwood(1980)]{Sell80}
Sellwood, J. A. 1980, \aap, {\bf 89}, 296

\bibitem[Sellwood(2006)]{Sell06}
Sellwood, J. A. 2006, \apj, {\bf 637}, 567

\bibitem[\protect\citeauthoryear{Sellwood}{2014}]{Se14}
Sellwood, J. A. 2014, arXiv:1406.6606 (on-line manual: \hfil\break {\tt http://www.physics.rutgers.edu/$\sim$sellwood/manual.pdf})

\bibitem[Sellwood(2016)]{Sell16}
Sellwood, J. A. 2016, \apj, {\bf 819}, 92

\bibitem[Sellwood \& Athanassoula(1986)]{SA86}
Sellwood, J. A. \& Athanassoula, E. 1986, \mnras, {\bf 221}, 195

\bibitem[Sellwood \& Carlberg(2014)]{SC14}
Sellwood, J. A. \& Carlberg, R. G. 2014, \apj, {\bf 785}, 137

\bibitem[Sellwood \& Kahn(1991)]{SK91}
Sellwood, J. A. \& Kahn, F. D. 1991, \mnras, {\bf 250}, 278

\bibitem[Sellwood \& McGaugh(2005)]{SM05}
Sellwood, J. A. \& McGaugh, S. S. 2005, \apj, {\bf 634}, 70

\bibitem[Toomre(1964)]{To64}
Toomre, A. 1964, \apj, {\bf 139}, 1217

\bibitem[Toomre(1966)]{To66}
Toomre, A. 1966, in {\it Geophysical Fluid Dynamics}, notes on the 1966 Summer Study Program at the Woods Hole Oceanographic Institution, ref. no. 66-46, p~111

\bibitem[Toomre(1981)]{To81}
Toomre, A. 1981, In ''The Structure and Evolution of Normal Galaxies'', eds.~S. M. Fall \& D. Lynden-Bell (Cambridge, Cambridge Univ. Press) p.~111

\bibitem[Tremaine \& Weinberg(1984)]{TW84}
Tremaine, S. \& Weinberg, M. D. 1984, \mnras, {\bf 209}, 729

\bibitem[Weinberg(1985)]{Wein85}
Weinberg, M. D. 1985, \mnras, {\bf 213}, 451

\end{thebibliography}
\end{document}